\documentclass{edm_article}

\usepackage{subfigure}
\usepackage{hyperref}
\usepackage[hyphenbreaks]{breakurl}

\begin{document}

\title{Prerequisite-driven Q-matrix Refinement for Learner Knowledge Assessment: A Case Study in Online Learning Context}

\numberofauthors{2}

\author{
\alignauthor
Wenbin Gan\\
        \affaddr{National Institute of Information and Communications Technology}\\
        \affaddr{Tokyo, Japan}\\
        \email{wenbingan@nict.go.jp}
\alignauthor
Yuan Sun\\
        \affaddr{National Institute of Informatics}\\
        \affaddr{ Tokyo, Japan}\\
        \email{yuan@nii.ac.jp}
}

\maketitle

\begin{abstract}

The ever growing abundance of learning traces in the online
learning platforms promises unique insights into the learner knowledge assessment (LKA), a fundamental personalized-tutoring
technique for enabling various further adaptive tutoring services in these  platforms. Precise assessment of learner knowledge requires the fine-grained Q-matrix, which is generally designed by experts to map the items to skills in the domain. 
Due to the subjective tendency, some
misspecifications may degrade the performance of LKA. 
Some efforts have been made to refine the small-scale Q-matrix, however, it is difficult to extend the scalability and apply these methods to the large-scale online learning context
with numerous items and massive skills. Moreover, the existing LKA models employ flexible deep learning models that excel at this task, but the adequacy of LKA is still challenged by the representation capability of the models on the quite sparse item-skill graph and the learners' exercise data. To overcome these issues, in this paper we propose a prerequisite-driven Q-matrix refinement framework for learner knowledge assessment (PQRLKA) in online context. We infer the prerequisites from learners' response data and use it to refine the expert-defined Q-matrix, which enables the interpretability and the scalability to apply it to the large-scale online learning context. 
Based on the refined Q-matrix, we propose a Metapath2Vec enhanced convolutional representation method to obtain the comprehensive representations of the items with rich information, and feed them to  
the PQRLKA model to finally assess the learners' knowledge. Experiments conducted on three real-world datasets demonstrate the capability of our model to infer the prerequisites for Q-matrix refinement, and also its superiority for the LKA task.

\end{abstract}

\keywords{Learner Knowledge Assessment, Prerequisite Inference, Q-matrix Refinement, Graph Representation Learning
} 

\section{Introduction}

The popularity of online learning has increased in recent years, with increasingly many
intelligent tutoring systems (ITSs) becoming available to learners \cite{hasanov2019survey,gan2020knowledge}. 
The application of techniques from artificial intelligence and
cognitive psychology has brought increasing intelligence to
these systems \cite{hasanov2019survey,gan2019ai}, and has led to a very wide range of
advantages, e.g., enabling the knowledge-acquisition through a series of individualized learning activities that accommodate learners with different needs and knowledge proficiencies \cite{Gan2020AIModeling,liu2021survey}. A key technique underlying these adaptive tutoring services is LKA \cite{Gan2020AIModeling,gan2022knowledgestructure,gan2022knowledge}. 

LKA assesses learner knowledge in the granularity of a set of specific skills, also called knowledge concepts, based on the learners' exercising logs in the systems, thereby providing the detailed information about their strength and weakness. 
Detailed assessment of skills requires a fine-grained mapping of items to skills, i.e., the Q-matrix \cite{desmarais2014refinement}. A binary Q-matrix indicates how the skills are combined to correctly answer the items, hence bridging the latent skill mastery patterns of learners and their explicit responses. 
However, building the Q-matrix for a domain is a non-trivial task \cite{desmarais2015combining,desmarais2013matrix}, which requires the involvement of domain experts. It is widely recognized that this subjective process may consist of some misspecifications \cite{chiu2013statistical,desmarais2014refinement}, which may negatively degrade the estimation of learner knowledge \cite{chen2017q}. 
To alleviate the subjective bias in manually-defined Q-matrix, existing researches in the data mining and cognitive diagnostics fields have proposed various methods to infer the Q-matrix in a data-driven manner \cite{liu2012data,sun2014alternating}, further refine the existing Q-matrix \cite{desmarais2014refinement,desmarais2015combining,chiu2013statistical,kang2019q}, and validate the Q-matrix by combining with the cognitive diagnostic models \cite{ma2020empirical,de2016general}. They also verified that in most of the cases, the refined Q-matrix has better model fit on the data than the original expert-labeled Q-matrix \cite{desmarais2013matrix,matsuda2015machine}. However, most of the efforts have been devoted to the scenarios of small tests with quite limited number of skills and items. Even for the famous fraction-subtraction dataset \cite{tatsuoka2013toward} with only twenty items and eight skills, the obtained Q-matrix is still controversial \cite{kang2019q}. It is difficult, and sometimes impractical, 
to extend the scalability and apply these methods to the large-scale online learning context with numerous items and massive skills.  

Moreover, most of the existing researches refine the Q-matrix by considering a set of Q-matrices in the latent space and determining the one with superior model fitness on the data using some fit strategies \cite{kang2019q}. Typical studies are the maxDiff \cite{de2008empirically}, minRSS \cite{chiu2013statistical}, and the ALS \cite{desmarais2013matrix}. These researches showed good performance on the learner data, but they generally ignore the explainable structural and inherent information among the questions and skills.   
Actually, the interdependencies between the
skills in a domain have long been acknowledged as \textit{prerequisites} in both cognitive and education science \cite{chen2018prerequisite}. 
In the real world, questions are interrelated to each other and are also closely related to the underlying skills required for their solution. When learners grow their knowledge from a certain question incorporating $skill_{1}$, they also improve their attainment of $skill_{2}$ to some extent (assuming that $skill_{1}$ is a prerequisite of $skill_{2}$). 
Correspondingly, a question with a specific skill also requires the prerequisite skills. 
These prerequisites between pedagogical concepts can be represented as a knowledge graph (KG) \cite{liu2019exploiting}, which contains additional structural information about the learning domain, and provides the potential for further refining the Q-matrix.  
Nevertheless, the prerequisites specifying the structure of skills has rarely been applied in the refinement of Q-matrix because obtaining the KG of a domain is labor-intensive \cite{pan2017prerequisite,zhang2016topological,nakagawa2019graph}.
Fortunately, the ever growing abundance of learning traces in the online learning platforms further enhances our capacity to obtain the prerequisites (KG) directly from the data through data mining techniques.
Based on this idea, we infer the prerequisites between skills from the data and use them to refine the Q-matrix, and apply the refined Q-matrix to assess the knowledge proficiency of learners. We try to balance the interpretability of Q-matrix refinement and the performance in applying it to the LKA task.

To assess learner knowledge in the online learning systems, plenty of researches have been conducted leveraging the compelling attribute of deep learning methods, such as the deep knowledge tracing (DKT) \cite{piech2015deep} and Dynamic key-value memory networks for KT (DKVMN) \cite{zhang2017dynamic}, and excelled at this task. 
However, most of these models used the underlying skills as input (also termed as skill-level KT models) and did not distinguish questions containing the same skills. The loss of distinctive information related to the questions may lead to imprecise inferences of the learners' knowledge states \cite{ghosh2020context,gan2022knowledgestructure}. To enrich the question representation, graph KT methods, such as the graph-based interaction KT (GIKT) \cite{yang2020gikt} and graph for KT
(GKT) \cite{nakagawa2019graph}, have been proposed to learn the embedding of questions from the question--skill graph. However, they ignore the
prerequisite information of the skills in the domain, and the sparseness of the original question--skill graph and the learner data greatly limit the capability of the graph learning methods. To overcome this sparseness, in this paper we refine the Q-matrix and propose a graph representation learning model to enrich the embedding learning from the prerequisite--enhanced dense question--skill graph and integrate it into the LKA task.

To summarize, In this paper we refine the Q-matrix for learner
knowledge assessment in online learning context, and propose a prerequisite-driven Q-matrix refinement framework for LKA (PQRLKA). Specifically, we first explore eight methods to infer the prerequisites from learners' response data in the online learning systems and use it to refine the expert-defined Q-matrix to eliminate the potential subjective tendency of experts in designing the Q-matrix.
This Q-matrix refinement method offers two advantages over the existing methods: first, it leverages the additional structural information about the learning domain to refine the Q-matrix rather than use the model fit indices to brutally change some entries from zeros to ones, thus enabling the interpretability of this process. Second, it refines the Q-matrix by applying the inferred prerequisites in a date-driven manner without replying on the specific cognitive diagnosis models (i.e., model-independent), and hence has good scalability. Based on the refined Q-matrix, we propose a graph representation learning model for LKA to verify its effectiveness in the online learning scenarios. A Metapath2Vec \cite{dong2017metapath2vec} enhanced convolutional representation method is then proposed to obtain the comprehensive representations of the attempted questions with rich information. 
These representations for the learners' exercising sequences are fed into the PQRLKA model, which considers the long-term dependencies using an attention mechanism, to finally predict the learners' performance on new problems. The main contributions are listed below.
\begin{itemize}
\item We explore eight methods that automatically discover the domain prerequisites from learner response data, and leverage it to refine the expert-defined Q-matrix. 
\item  We propose the PQRLKA framework to assess the learner knowledge with a Metapath2Vec enhanced convolutional representation method for  comprehensive question representations with rich information.
\item  We integrate the Q-matrix refinement with the task of learner knowledge assessment, joint these two tasks together and leverage the former to benefit the later.
\item We conduct comprehensive experimental evaluations on three real-world datasets. The results demonstrate the superiority and effectiveness of our method in Q-matrix refinement and learner knowledge assessment in online learning context. 
\end{itemize}

\section{Related Work}

\subsection{Q-matrix Refinement}

Due to the well-acknowledged issue of subjective tendency of expert-defined Q-matrix, a fair number of researches have been conducted to refine the Q-matrix based on the empirical data. 
Some studies have investigated the influence of the misspecified Q-matrix in learner knowledge assessment. Chen \cite{chen2017q} studied the effects of two types of Q-matrix misspecification (Q-entry and attribute misspecification), and found that both types can significantly deteriorate the classification accuracy and consistency of diagnostic results.

To obtain an optimal Q-matrix, existing efforts can be classified into two main categories: Q-matrix refinement and Q-matrix derivation \cite{minn2016refinement}. 
Q-matrix refinement optimized the Q-matrix from a given plausible one while the Q-matrix derivation infer the full Q-matrix from the beginning. Mathematically, these two problems can be formulated as finding a proper function $f$ that $Q_{refinement}=f(R, Q_{expert})$ and $Q_{derivation}=f(R)$, where $R$ is the response data. In Q-matrix refinement, Typical studies are the Maximum Difference (MaxDiff) \cite{de2008empirically}, Minimal Residual Sum Square (MinRSS) \cite{chiu2013statistical}, and the Conjunctive Alternating Least Square Factorization (ALSC) \cite{desmarais2013matrix}.
These researches defined different strategies to select the optimal Q-matrix from learner data, and were proved to be effective on this task. 
Researchers also proposed superior model fit indices to determine the optimal Q-matrix. Kang et al. \cite{kang2019q} 
proposed an item fit statistic root mean square error approximation for validating a Q-matrix with the deterministic inputs, noisy, ``and'' (DINA) model. De la Torre \cite{de2016general} proposed a discrimination index that can be used with a wide class of CDM to empirically refine the Q-matrix. To more effectively refine the Q-matrix, 
Xu et al. \cite{xu2016boosted} adopted ensemble algorithms and combined the boosting technique with a decision tree to fuse the suggested refinements of three existing methods. Their results yielded substantial gains over the best performance of individual method. 
Most of these methods are limited to data with small numbers of items and skills, 
when the number of items or skills becomes large, the computational burden for optimizing the Q-matrix grows significantly \cite{chen2017q}.
Hence it is difficult, and even impractical, to directly apply these methods to the large-scale online learning context. 

Some researchers explored to directly derive the optimal Q-matrix from the response data. 
Matrix factorization (MF) techniques, such as non-negative MF \cite{desmarais2012mapping} and Boolean MF \cite{sun2014alternating}, are popular methods to derive Q-matrix from the data.
Liu et al. \cite{liu2012data} proposed a data-driven approach to identification of the Q-matrix and estimation of related model parameters in DINA model. These methods do not reply on the expert-defined Q-matrix completely, and the derived Q-matrices from data often show better predicting performance when jointing with the specific CDM models. However, this line of work is controversial as some researchers pointed out that the expert-defined Q-matrix should not be overlooked in the refinement process as the results based only on statistical information might be misleading \cite{chen2017q}. In addition, the deviation methods suffer from the issue of interpretability, as it is difficilt to explain the item-skill mappings generated from the data. In this paper, we derive the Q-matrix from the learner response data by considering the experts' opinion simultaneously, and try to balance the interpretability of Q-matrix refinement and the performance in applying it to the online learning context.

\vspace{- 10 pt}
\subsection{Prerequisite Inference}
Prerequisites are generally defined by the experts, several studies have been attempted to automate this process. 
Gasparetti \cite{gasparetti2022discovering} proposed an approach for prerequisite discovering in educational documents based on word embedding. It represented the text-based
learning objects in low-dimensional latent spaces and
identified prerequisites in a binary classification setting.
Sabnis et al. \cite{sabnis2021upreg} used NLP techniques to automatically infer prerequisite relationships between different concepts in the online courses. They statistically compared the relevancy scores between a
pair of concepts with high semantic similarity to determine the prerequisite.
The above methods are ambiguous on the set of skills and 
infer the prerequisite using NLP analysis on the large corpus of educational documents, which may not suitable for the exercising logs with no textual information. 

Wang et al. \cite{wang2021learning} proposed a latent-variable selection method with regularization for cognitive diagnostics, which learns the skill hierarchies from learner response data. Using a DKT model, Zhang et al. \cite{zhang2016topological} discovered the topological order of skills from learners' exercise performance. Like our group, they intuited that the KG is difficult to directly extract from response data, but this difficulty can be bridged by ordering the learners' mastery of skills. Hence, in this paper, we discover the prerequisite skill pairs in the KG by ordering the learners' performances in question-solving (i.e., their mastery of the underlying skills). 

\begin{figure*}[tbp]
\centering
\includegraphics[width=1.\textwidth]{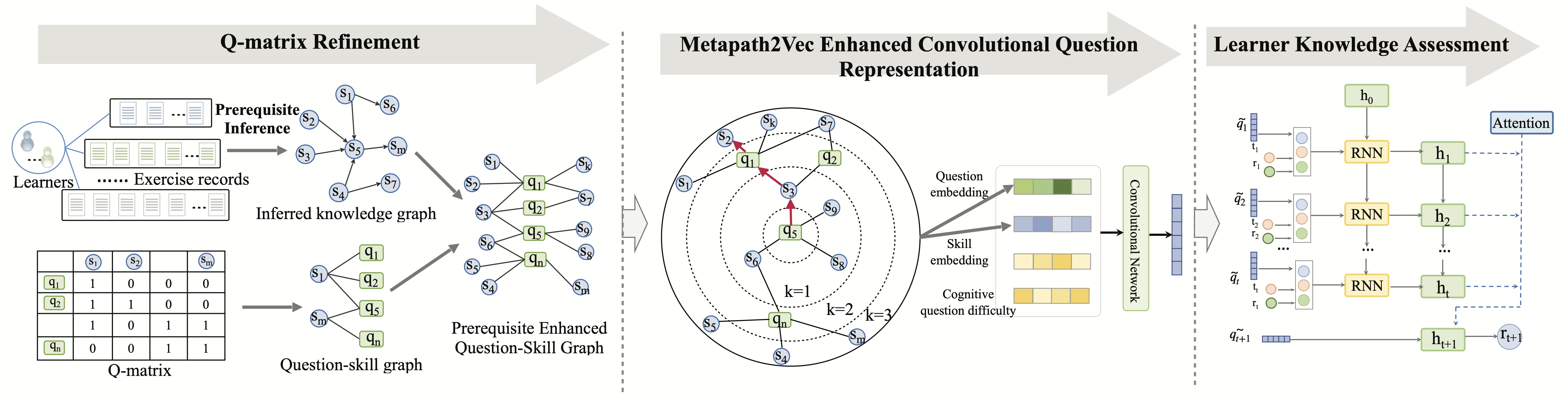} 
\caption{The framework of the proposed method.}
\label{framework}
\end{figure*}

\vspace{- 5 pt}
\subsection{Learner Knowledge Assessment}
LKA is to obtain learner knowledge states based on the learners' explicit learning feedback on various types of learning materials, such as the exercises and video lectures \cite{liu2021survey,Gan2020AIModeling,gan2022knowledge}. In this paper, we focus on the learner knowledge assessment by utilizing their performance on exercises, which can be interpreted as an explicit indication of their implicit knowledge. Based on the application context, existing work can be divided into two categories: static LKA for testing and dynamic LKA for learning. 

\vspace{- 3 pt}
Static LKA, also termed as Cognitive Diagnostic Assessment (CDA), is usually conducted after a summative test through some diagnostic models. This topic has been widely explored by researchers in the fields of psychometrics and data mining. IRT \cite{baker2001basics} and DINA \cite{de2009dina} are two of the most renowned models in this category. IRT diagnoses a learner's knowledge proficiency using a unidimensional variable (i.e., latent trait). The DINA model \cite{de2009dina} describes each learner's proficiency level using a binary vector. Unlike IRT, the DINA model must infer this multidimensional vector using the Q-matrix. 
Meanwhile, data mining researchers mainly conduct this task using Matrix factorization (MF) \cite{desmarais2012mapping,sun2014alternating}. 
As these methods do not consider the temporal factor of learning, it is difficult to be directly applied to the online learning context.

\vspace{- 3 pt}
Dynamic LKA, also known as Knowledge Tracing (KT), models the learner performance from a long period and traces the learners' knowledge over time utilizing the temporal factor of learning in ITSs \cite{piech2015deep,zhang2017dynamic,huang2020learning,vie2019knowledge,jiangimproving2021}. 
The existing KT methods can be generally divided into three main categories: probabilistic models, factor analysis models and the deep learning models \cite{liu2021survey,gan2022knowledge}. 
Bayesian knowledge tracing (BKT) \cite{corbett1994knowledge}, a typical probabilistic model, used the hidden Markov model to separately track the learner proficiency on each skill. Factor-analysis models, such as the additive factor model (AFM) \cite{cen2006learning} and performance factor analysis (PFA) \cite{pavlik2009performance}, pre-define a delicate model framework and assign the considered factors as model parameters, which are learned from the data to generalize the observations. A famous study employed a knowledge tracing machine (KTM) based on the factorization machine, providing a generic framework for incorporating learning-related side information into a student model \cite{vie2019knowledge}. Further researches was conducted recently based on the KTM, for example the DAS3H (item Difficulty, student Ability, Skill, and Student Skill practice History) \cite{choffin2019das3h} and RKTM (recurrent KTM) \cite{lai2021recurrent} by enriching the KTM with additional information. 
Deep learning based models are now popularly used for the KT task and show significant improvements in model performance over traditional models. Typical researches in this direction are DKT \cite{piech2015deep}, DKVMN \cite{zhang2017dynamic} and AKT \cite{ghosh2020context}. 
However, in most of these models, the KT is based on the skills in a specific domain. Such skill-level KT models do not distinguish questions containing the same skills. The loss of distinctive information related to the questions may lead to imprecise inferences of the learners' knowledge states. Moreover, almost all of these models simply assume that all questions and skills are independent, which is unrealistic in the actual learning process.

\vspace{- 3 pt}
The relations between questions and skills have been considered in various graph-based KT models \cite{liu2020improving,yang2020gikt,tong2020hgkt,nakagawa2019graph}. Liu et al. \cite{liu2020improving} built a question--skill bipartite graph based on the Q-matrix in the domain and pre-trained the question and skill embedding in the graph using three constraints of explicit question--skill relations and implicit relations of skill similarity and question similarity. As the dense embeddings of questions and skills contain the relations in the graph, the pre-trained embedding-fed KT model outperforms the baseline models. Yang et al. \cite{yang2020gikt} proposed the GIKT model and incorporated the question--skill correlations via embedding propagation on the question--skill relation graph using a graph convolutional network (GCN). 
Tong et al. \cite{tong2020hgkt} introduced problem schema with a hierarchical exercise graph to KT. The problem schema clustered similar exercises into the same group to incorporate the question--question relations. Pandey et al. \cite{pandey2020rkt} calculated the exercise--exercise relations based on the performance data and the exercise text, and incorporated it into a transformer model for relation-aware KT. 
All of these methods exploit the relation information in various graphs and introduce the distinctive information of questions into the KT task. However, they ignore the prerequisite information of the skills in the domain. 

\vspace{- 3 pt}
Similarly to our work, Nakagawa et al. \cite{nakagawa2019graph} and Chen et al. \cite{chen2018prerequisite} incorporated the KG information into the KT procedure. Chen et al.'s work assumes that the KG is already given by experts, and models the prerequisite as an ordering pair. In this way, the mastery of related skills is constrained by referring to the KG. Nakagawa et al. used the KG
information and a graph neural network (GNN) to update the learners' hidden knowledge states. Unlike these earlier works, our method learns the prerequisites from learner response data to build the KG automatically, and integrates it to refine the Q-matrix to build a KG-enhanced relation graph. Leveraging the Metapath2Vec method, we learn the dense embeddings of all nodes, which incorporate the multi-hop question--question, question-skill, and skill--skill relations in the domain.

\vspace{- 15 pt}
\section{Proposed Method}

This section introduces the proposed PQRLKA framework for assessing learner knowledge, as shown in Figure \ref{framework}. The framework proceeds the LKA task using three modules: Q-matrix refinement, Metapath2Vec enhanced convolutional question 
representation, and learner knowledge state evolution.
It first builds a KG for the learning domain by inferring the prerequisite relations from the learner response data, and then uses it to refine the expert-defined Q-matrix, thus generating a refined Q-matrix, which can be naturally represented as a 
prerequisite enhanced question--skill graph. 
Based on this enhanced graph, it conducts the question representation learning based on a Metapath2Vec enhanced convolutional neural network. Specifically, the dense embeddings of all skill and question nodes are obtained through the graph embedding learning method Methpath2Vec. To incorporate more distinctive information of the questions, the cognitive question difficulty at both the question and skill levels are also inferred from the learning histories of individual learners. A convolutional representation method is then proposed to fuse the question and skill embeddings with the cognitive question difficulty. It considers each separate factor and the interactions between each pair of factors, thus obtaining the comprehensive representations of questions (see Figure \ref{fig3}).
Compared with the existing methods, the proposed question representation learning method incorporates both the distinctive information of the specific question and the various relations between questions and skills, a factor making it more superior for the later assessment of learner knowledge. 
These representations are then fed into the attentive LKA network for predicting learner performance. It is worth noting that we conduct the Q-matrix refinement and also the learner knowledge assessment together.

\vspace{- 7 pt}
\subsection{Q-matrix Refinement}

\subsubsection{Prerequisite Inference from Data}
In practical educational scenarios, there always exists a topological order (prerequisites) among the skills in a domain, because skills are taught and learned in sequence. In many learning systems the KG information is never provided and must be obtained by time-consuming labor. We observed that if learners have not mastered a specific skill, their probability of incorrectly answering questions requiring the post-requisite skills will increase. Moreover, learners' mastery of a skill can be indicated by their performances on the attempted questions requiring that skill. Based on these observations, this paper aims to discover the prerequisites and built the KG from the learner response data.

Here we infer the prerequisites from the order of the learners' mastery of skills, which is explicitly represented by the exercising performance data. We first build a skill relation matrix $R \in \mathbb{R}^{|S|\times |S|}$, in which entry $R_{i,j}$ represents the prerequisite relation $s_{i} \rightarrowtail s_{j}$ between skill $s_{i}$ and $s_{j}$. Inspired by the definition of question similarity in previous methods \cite{nazaretsky2019kappa,tong2020hgkt,pelanek2019measuring}, we explore the underlying KG using eight methods: Skill Transition, Cohen’s Kappa, Adjusted Kappa, Phi coefficient, Yule coefficient, Ochiai coefficient, Sokal coefficient, Jaccard coefficient, as described below.

\vspace{-3 pt}
\textbf{Skill Transition}: The skill-transition matrix $R$ contains the transitions of different skills. Its entries are 
$R_{i,j}^{SK} = \frac{n_{i,j}}{\sum_{k=1}^{|\kappa|}n_{i,k}}$, where $n_{i,j}$ denotes the number of times in which skill $s_{j}$ is trained immediately after training skill $s_{i}$.   

\begin{table}[t]
\renewcommand{\arraystretch}{1.0}
\centering
\caption{Contingency table for a pair of skills $s_{i}$ and $s_{j}$}
\label{tab2}
\resizebox{0.8\linewidth}{!}{%
\begin{tabular}{c|ccc}
\hline
             & $\boldsymbol{s_{j}}$ master &  $\boldsymbol{s_{j}}$ not master & total   \\ \hline
$\boldsymbol{s_{i}}$ master   & a          & b            & a+b     \\
$\boldsymbol{s_{i}}$ not master & c          & d            & c+d     \\ 
total       & a+c        & b+d          & a+b+c+d \\ \hline
\end{tabular}}
\end{table}
\begin{table}[tbp]
\renewcommand{\arraystretch}{1.2}
\centering
\caption{Evaluation indices for obtaining KG using  the contingency table}
\label{evaluation_indices}
\resizebox{1.\linewidth}{!}{
\begin{tabular}{lc} 
\hline
Evaluation Index & Formula \\ \hline
{Cohen’s Kappa} &
$R_{i,j}^{Kappa} = {2(ad-bc)}/{\lbrace(a+b)(b+d)+(a+c)(c+d)\rbrace}$ \\ 
{Adjusted Kappa} &
$R_{i,j}^{Kappa'} = {2(ad-bc)}/{\lbrace(a+c)(c+d)\rbrace}$ \\
{Phi coefficient}&
$R_{i,j}^{Phi} = {(ad-bc)}/{\sqrt{(a+b)(b+d)(a+c)(c+d)}}$ \\
{Yule coefficient}&
$R_{i,j}^{Yule} = {(ad-bc)}/{(ad+bc)}$ \\
{Ochiai coefficient}&
$R_{i,j}^{Ochiai} = {a}/{\sqrt{(a+b)(a+c)}}$ \\
{Sokal coefficient}&
$R_{i,j}^{Sokal} = {(a+d)}/{(a+b+c+d)}$ \\
{Jaccard coefficient}&
$R_{i,j}^{Jaccard} = {a}/{(a+b+c)}$ \\
\hline  
\end{tabular}}
\end{table}

\vspace{-1 pt}
To further leverage the impact of the learners' performance of one skill on the performance of another, we summarized the learners' performance on skill pair $s_{i}$ and $s_{j}$ in a contingency table (see Table \ref{tab2}). As mentioned above, we interpreted the learners' correct or incorrect responses as mastery indicators of the underlying skills of the given questions. It is worth noting that in Table \ref{tab2}, the question requiring skill $s_{i}$ occurs before the question requiring $s_{j}$ in the learning sequence. When there are multiple occurrences of question pairs in the learning sequence, we consider only the latest occurrence. Based on the contingency table, we discovered the KG using the evaluation indices \cite{pelanek2019measuring} in Table \ref{evaluation_indices}, which measure the agreement of the prerequisite relation between a pair of skills. These indices are widely used for measuring associations between two variables.

\vspace{-3 pt}
As the KG is always a unidirectional graph, we simplified it by a suitable strategy. We also imposed a threshold that controlled the sparsity of the relations in KG. The final skill relation matrix was denoted as $R^{w}$, $w \in \lbrace SK, Kappa,Kappa'$,\\$Phi,Yule,Ochiai,Sokal,Jaccard \rbrace$. The elements along the diagonal of $R^{w}$ were set to one.
\begin{equation}  
\begin{cases}
R_{i,j}^{w} = max(R_{i,j}^{w},  R_{j,i}^{w}), R_{j,i}^{w} = 0, & if R_{i,j}^{w} \geq  R_{j,i}^{w} \\
R_{j,i}^{w} = max(R_{i,j}^{w},  R_{j,i}^{w}), R_{i,j}^{w} = 0, & otherwise
\end{cases}
\end{equation}
\begin{equation}  
R_{i,j}^{w} =
\begin{cases}
1, & if  R_{i,j}^{w} \geq threshold \\ 
0, & otherwise
\end{cases}
\end{equation}

\subsubsection{Refine the Expert-defined Q-matrix}

\begin{figure}[tbp]
\centering
\includegraphics[width=1.\linewidth]{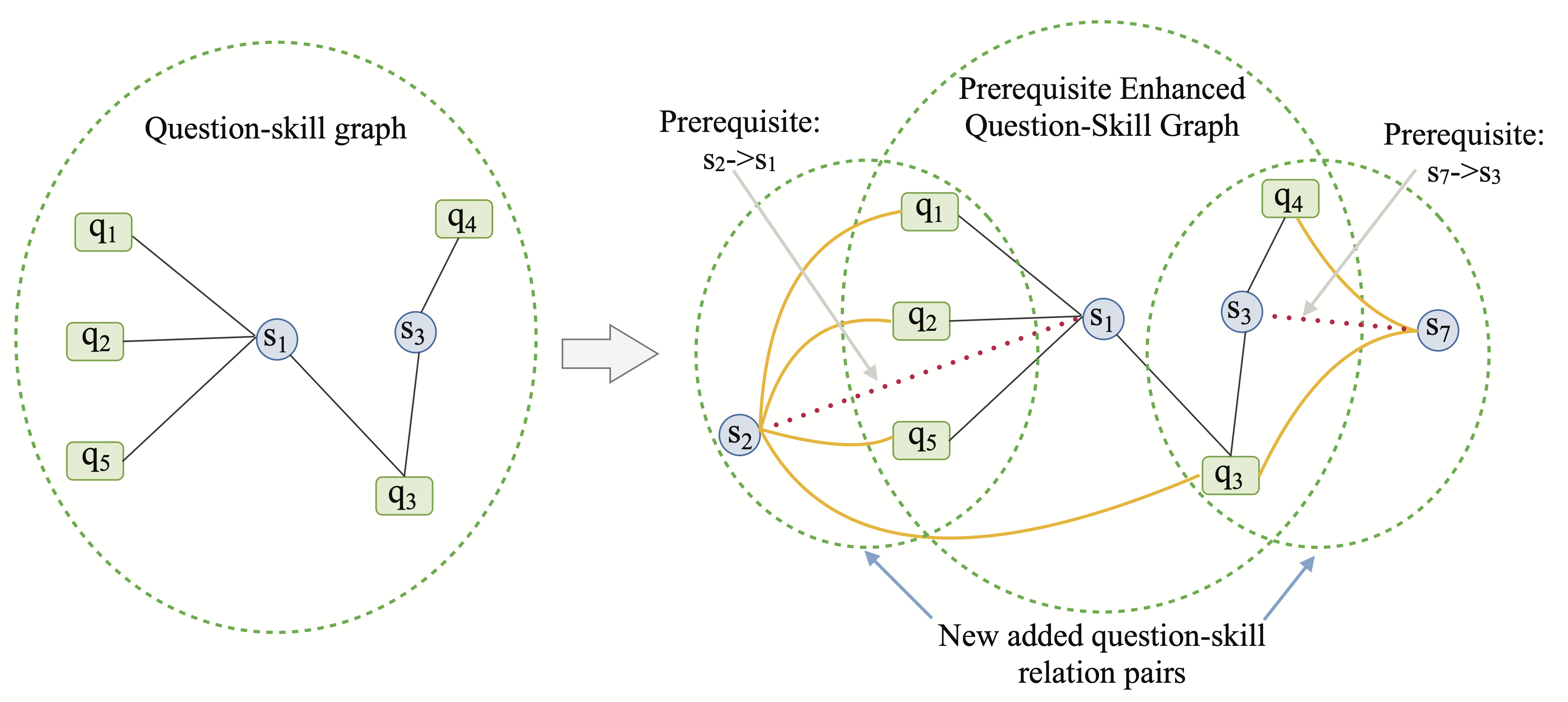} 
\caption{The procedure of Q-matrix refinement.}
\label{Q_matrix_refinement}
\end{figure}

After obtaining the inferred skill relation matrix $R^{w}\in \mathbb{R}^{|S|\times |S|}$ of the KG that includes all the prerequisites among skills, we refine the original expert-defined q-matrix $O\in \mathbb{R}^{|Q|\times |S|}$.

The matrix representation $\hat{O}\in \mathbb{R}^{|Q|\times |S|}$ of the refined new Q-matrix is obtained as follows:
\begin{equation}  
\hat{O} = O (R^{w})^{T}
\end{equation}
\begin{equation}  
\hat{O}_{i,j} =
\begin{cases}
1, & if  \hat{O}_{i,j} \geq 1 \\ 
0, & otherwise
\end{cases}
\end{equation}
In Eq. (3), the transpose of the skill relation matrix $R^{w}$ accounts for the skills that are prerequisite to the skills of the current question. In other words, a question requiring a specific skill is also related to the prerequisite skills. The procedure of the refinement can be shown in Figure \ref{Q_matrix_refinement}.

\subsection{Learner Knowledge Assessment Based on Refined Q-matrix}

\subsubsection{Metapath2Vec Enhanced Convolutional Question Representation}

The matrix $\hat{O}$ can be naturally represented as a graph, hereafter denoted as the KG-enhanced question--skill graph. This graph includes not only the multi-hop relations between questions and skills, but also the prerequisite relations among the skills. The embeddings in this graph were obtained using the adapted Metapath2Vec method \cite{dong2017metapath2vec}.

The adapted Metapath2Vec method proceeds in two main steps: meta-path generation and skip-gram-based embedding learning. A meta-path is a sequence of nodes following the edges in the graph. To assure that all questions and skills appear in the final embeddings, we generate the meta-paths from the KG-enhanced question--skill graph using a question--skill--question (QSQ) pattern, in which every meta-path begins with a question node followed by a skill node and then by a question node; for example, $\rho$: $q_{1}\stackrel{\hat{O}_{1,1}}{\longmapsto}s_{1}\stackrel{\hat{O}_{2,1}}{\longmapsto}q_{2}\stackrel{\hat{O}_{2,2}}{\longmapsto}s_{2}\stackrel{\hat{O}_{3,2}}{\longmapsto}q_{3}$. Setting two hyper-parameters---the path length $\wp $ and number of paths $\aleph $---for each question node, we generated all meta-paths on the graph.
The heterogeneous skip-gram is leveraged to learn the node embeddings by maximizing the probability of having context $N_{t}(v_{t})$ given a node $v_{t}$, where $N_{t}(v_{t})$ is the set of one-hop neighbors of nodes $v_{t}$ following the QSQ pattern. The interested reader can refer to 
\cite{dong2017metapath2vec} for the original Metapath2Vec method.
Finally, we obtain all embeddings ($q$ for questions and $s$ for skills) with the same dimension $d$ of question and skill nodes in the graph, which also enclose the relation information.  

\begin{figure}[tbp]
\centering
\includegraphics[width=0.5\textwidth]{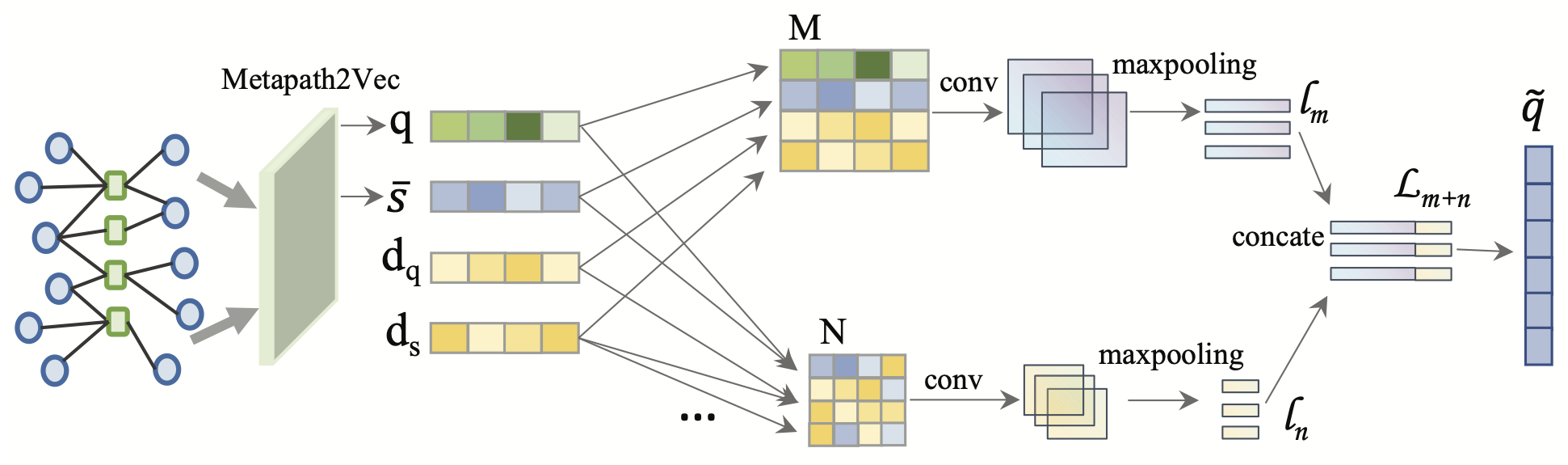} 
\caption{Use of Metapath2Vec enhanced convolution to fuse various distinctive features and their interactions into the comprehensive question representation}
\label{fig3}
\end{figure}
\vspace{-5 pt}

To distinctively represent the questions, 
we fuse the various feature representations through convolutional operations to obtain the comprehensive question embeddings with additional distinctive information, i.e., the skill information and the cognitive difficulty information \cite{gan2019field}. The cognitive difficulty is calculated following prior work in \cite{Gan2020AIModeling}.
Following \cite{liu2020improving} and \cite{yang2020deep} that learn the high-order latent patterns through feature interactions and convolution operations (rather than directly concatenating the features), we map and fuse the separate features and their interactions using convolution operations, as shown in Figure \ref{fig3}.

The question-difficulty information is represented as vectors using an embedding matrix $D$ (of size $(c+1) \times d$). The continuous embedding vectors at the question and skill levels of each question are defined as $d_{q}=\Psi_{q,t}D $ and $d_{s}=\Psi_{s,t}D$, respectively (referring $\Psi_{q,t}$ and $\Psi_{s,t}$ in \cite{Gan2020AIModeling}). For questions containing multiple skills, we represent the skill as the average skill embedding $\overline{s} = Avg(s_{1},...,s_{q})$.
Fusing the above-obtained features, we generate the linear information $M$ and quadratic information $N$ for question $q$. 
\begin{equation}  
\begin{aligned}
M &= [q,\overline{s},d_{q},d_{s}]\in \mathbb{R}^{4\times d},\\
N &= [\langle M_{i},M_{j} \rangle]\in \mathbb{R}^{4\times 4},
\end{aligned}
\end{equation}
where $\langle.\rangle$ represents the interactions of two vectors obtained by the inner product. We then apply the two-dimensional convolution operation with eight kernels of size $2\times2$ on both $M$ and $N$, and maxpooling on each feature map to obtain $l_{m}\in \mathbb{R}^{1\times(d-1)\times 8}$ and $l_{n}\in \mathbb{R}^{1\times3\times 8}$. These two parts are then concatenated into eight longer vectors $l_{m+n}\in \mathbb{R}^{1\times(d+2)\times 8}$ including the convolutions from the separate features and their interactions. Inspired by the multi-head mechanism in the transformer model \cite{vaswani2017attention}, we concatenate and linearly transform the eight vectors and hence obtain the final question representation $\widetilde{q}\in \mathbb{R}^{1\times d^{'}}$. 
\begin{equation}  
\begin{aligned}
&l_{m} =  MaxPooling(Conv(M)),\\
&l_{n} =  MaxPooling(Conv(N)),\\
&l_{m+n} =  Concat(l_{m},l_{n}),
\end{aligned}
\end{equation}
\begin{equation}  
\widetilde{q} = Concat(l_{m+n}^{1},...,l_{m+n}^{8})W^{O}
\end{equation}
where $W_{O}\in \mathbb{R}^{((d+2)\times 8)\times d^{'}}$ is the parameter that transforms the convolution results into a vector. 

\vspace{-5 pt}
\subsubsection{Learner Knowledge State Evolution}
Since the dynamic evolution of learner knowledge in the online learning systems, the learner exercising sequences are fed into an attentive KT framework that predicts the learner performance, as shown in the right part of Figure \ref{framework}.

The log data of each interaction in the exercising sequences consists of a tuple representing the question, the correctness, and the elapsed time. Look-up operations are performed on an embedding matrix $E_{r}\in \mathbb{R}^{2\times d^{'}}$, in which row vector $r_{t}$ contains the incorrectness or correctness of the responses. The elapsed time $et$ strongly evidences a student's proficiency in knowledge and skills \cite{shin2020saint+}. This time is converted to seconds and capped at 500 seconds. A $d^{'}$-dimensional latent embedding vector for $et_{k}$ is computed as $t_{k}= et_{k}W_{et}+b_{et}$, where $W_{et}$ and $b_{et}$ are learnable vectors. The interaction embedding is obtained as 
$  x_{t}= Concat(\widetilde{q},t_{t},r_{t}).$


The sequence data of the learners' exercising process are modeled using LSTM \cite{piech2015deep} within the KT framework, and the learners' knowledge states are 
traced at each time. Since KT problem can be regarded as a prediction problem based on the long-term time-series data, LSTM is especially suitable for such kind of task and can extract useful information in the previous learning histories to predict future performance. 
Here we adopt the LSTM in the final part of our framework is also for the fair consideration, to better compare with other KT methods and show the effectiveness of the Q-matrix refinement for this task. 
The hidden knowledge state $h_{t}$ of a learner at step $t$ is updated based on the current input and the previous state as:
\begin{equation} 
h_{t}=LSTM(x_{t},h_{t-1}; \theta)
\end{equation}
We then employ an attention mechanism that accounts for the impact of previous attempts on the current attempt. A new question will likely be strongly affected by similar questions or questions requiring the same skillset as the new question. Moreover, according to forgetting-curve theory, the impact exponentially decays as time passes \cite{ghosh2020context,Gan2020AIModeling}. To describe these effects, we assume that the learner-knowledge state in the current step is the weighted sum of the aggregated states in the previous steps. The weights are based on the correlations: 
\begin{equation}
h_{t+1}= \sum_{i=1}^{t}\alpha_{i,t+1} h_{i}.
\end{equation}
The attention was calculated using a combination of the shared skill-based attention and the question similarity-based attention, as below.
\begin{equation}
\begin{aligned}
&\alpha_{i,t+1} = \frac{exp(corre_{i,t+1})}{\sum_{j}exp(corre_{j,t+1})},\\
&corre_{i,t+1} = exp(-\theta |t_{t+1}-t_{i}|) g(\widetilde{q_{i}},\widetilde{q}_{t+1}),\\
\end{aligned}
\end{equation}
where $|t_{t+1}-t_{i}|$ is the time interval between step $i$ and $t+1$, $\theta > 0$ is the learnable decay rate over time, the exponential term down-weights the importance of questions in the distant past, and $g(.)$ denotes the correlation between two questions.
\begin{equation}
\begin{aligned}
& g(\widetilde{q_{i}},\widetilde{q}_{t+1}) = \lambda g_{1}(\widetilde{q_{i}},\widetilde{q}_{t+1}) + (1-\lambda)g_{2}(\widetilde{q_{i}},\widetilde{q}_{t+1})\\
& g_{1}(\widetilde{q_{i}},\widetilde{q}_{t+1}) = \frac{n}{|s_{q_{t+1}}|}\\
& g_{2}(\widetilde{q_{i}},\widetilde{q}_{t+1}) = cos(\widetilde{q_{i}},\widetilde{q}_{t+1})
\end{aligned}
\end{equation}
where $\lambda$ is a tunable parameter that balances the above two correlations, and is set to 0.5 in our experiment. $n$ in $g_{1}$ is the number of shared skills and $|s_{q_{t+1}}|$ denotes the number of skills contained in question $q_{t+1}$. The cosine similarity in $g_{2}$ is calculated using the two vectors of question representations.

The learner performance at step $t+1$ can be  predicted from the question representation $\widetilde{q}_{t+1}$ and the current knowledge state $h_{t+1}$ as follows:
\begin{equation}
\begin{aligned}
&s_{t+1} = tanh(W_{s}[\widetilde{q}_{t+1},h_{t+1}])+b_{s},\\
&p_{t+1} = \sigma(W_{p}s_{t+1}+b_{p}),\\
\end{aligned}
\end{equation}
where $W_{*}$ and $b_{*}$ are parameters in the fully connected layer and the sigmoid activation layer, respectively.

Finally, we optimized our model by the cross-entropy loss; specifically, we minimized the following objective function between the true answer $ l_{t} $ and the predicted performance $ p_{t+1} $ at each interaction:  
\begin{equation}
\mathcal{L} =  -\sum_{t} (r_{t+1}log \, p_{t+1}+(1-r_{t+1})log (1-p_{t+1})).
\end{equation}

\section{Experiments}

\subsection{Datesets}
Experiments were performed on three well-established datasets, namely, Assist0910\footnote{Assist0910:\url{https://sites.google.com/site/assistmentsdata/home/assistment-2009-2010-data/skill-builder-data-2009-2010}}, Assist1213\footnote{Assist1213:\url{https://sites.google.com/site/assistmentsdata/home/2012-13-school-data-with-affect}}, and EdNet\footnote{EdNet:\url{http://bit.ly/ednet_kt1}}. All datasets are public real-world datasets containing the temporal interaction records between learners and real computer-aided tutoring systems. 

\vspace{-5 pt}
\textbf{Assist0910:} This dataset was collected from 2009 to 2010 using the ASSISTment system \cite{feng2009addressing}. It contains the interactions of 3002 learners and 17,705 items associated with 123 skills, forming 277,540 interaction records. 
\vspace{-5 pt}

\textbf{Assist1213:} This dataset, which updates Assist0910, was collected over the whole year of 2012. In this dataset, 22,339 learners in the system attempted 52,825 mathematics questions requiring 265 skills. This dataset contains the largest number of learners, questions, and interaction entries (nearly 2.7 million). It is a one-skill-per-question dataset, meaning that each question requires one skill. 
\vspace{-5 pt}

\textbf{EdNet:} The EdNet dataset is newly available to the public \cite{choi2020ednet}. Following \cite{yang2020gikt}, we randomly selected 5000 students who answered 12,372 questions requiring 188 skills, thus obtaining 347,866 interaction logs. This dataset contains the minimum number of attempted items per learner (69.57) but the largest number of skills per item among the three datasets. 

\vspace{-10 pt}
As done in existing studies \cite{vie2019knowledge,yang2020gikt}, we eliminated noise by deleting users with fewer than 10 interaction entries and questions with ``not-a number'' (NaN) skills from the three datasets. Table \ref{Datasets} summarizes the statistics of the datasets.

\begin{table}[t]
\centering
\caption{Statistics of the three datasets used in this study}
\label{Datasets}
\resizebox{1\linewidth}{!}{%
\begin{tabular}{llcr}
\hline
{Dataset}                           & {Assist0910} & {Assist1213} & {Ednet} \\ \hline
\# of learners                    & 3,002         & 22,339       &  5,000                    \\
\# of questions                   & 17,705     & 52,825     & 12,372                    \\
\# of skills                      & 123         & 265        & 188                       \\
\# of interaction records                & 277,540     & 2,672,532    & 347,864                 \\
\# skills per item                & 1.20       & 1.00          & 2.28                         \\
\# questions per skill            & 172.33           & 199.34    &149.78                        \\
\# of attempted items per learner & 92.45     & 119.64     &  69.57                         \\ \hline
\end{tabular}}
\end{table}

\subsection{Compared Models}
As part of our model evaluation, we competed the model against several state-of-the-art skill-- and question--based KT models. The performances of the various models were compared.
\vspace{-15 pt}
\begin{itemize}
\item \textbf{BKT} \cite{corbett1994knowledge} conducts KT using a hidden Markov model, and represents the learner knowledge states as a set of binary variables.
\vspace{-5 pt}
\item \textbf{DKT} \cite{piech2015deep} is the first KT model based on a deep neural network. This model treats the learner-knowledge prediction as a sequence learning task and captures the complex representations of student knowledge using the hidden vectors of RNNs. The input is a one-hot encoding of skills. 
\vspace{-5 pt}
\item \textbf{DKVMN} \cite{zhang2017dynamic} establishes the learners' knowledge states using an auxiliary memory that augments the neural networks. This method embeds the skill information into a key matrix and accumulates the temporal information from the learners' exercising sequences. It then infers their knowledge states on these skills. 
\vspace{-5 pt}
\item \textbf{KTM} \cite{vie2019knowledge} is a factor analysis model based on the factorization machine. It is a generic framework that incorporates the side information into the student model.   Learner performance is predicted based on a sparse set of weights applied to all features in the samples.
\vspace{-5 pt}
\item \textbf{DKT-Q} is a variant of DKT that replaces the skills embedding with a one-hot encoding of questions as the input.
\vspace{-5 pt}
\item \textbf{DKT-Q\&S} is a variant of DKT that inputs both the questions and skill representations to the DKT. 
\vspace{-5 pt}
\item \textbf{DKT-CQE} is a variant of DKT that inputs our convolutional question embeddings to the DKT. 
\vspace{-5 pt}
\item \textbf{GIKT} \cite{yang2020gikt} is a graph-based interaction model that learns the question representations from the question--skill relation graph using a GCN and conducts KT within the LSTM framework.
\vspace{-5 pt}
\item \textbf{DAS3H} \cite{choffin2019das3h} is a factor analysis model built on the KTM, it considers item difficulty, student ability, skill, and student skill practice history.  
\vspace{-5 pt}
\item \textbf{RKTM} \cite{lai2021recurrent} is a a recurrent knowledge tracing machine model by temporally enriching the
encoding of KTM and DAS3H with the knowledge state of students.
\vspace{-5 pt}
\item \textbf{AKT} \cite{ghosh2020context} is an attentive KT model, it builds context-aware representations of questions
and utilizes a novel monotonic attention mechanism to relate a learner's
current answers to questions to their past responses.

\end{itemize}
\vspace{-15 pt}
Among the baseline models, the former three are skill-based models, in which the LKA is based on the skills contained in the questions. The latter eight and the proposed model are question-based models that account for the distinctive question information.

\subsection{Setup and Implementation}

Before conducting the experiments, we extracted 20\% of the sequences in the dataset as the test set and retained the remaining 80\% as the training set.

\vspace{-3 pt}
To embed the nodes in the graph using Metapath2Vec, we set the length of all meta-paths as $\wp=7 $ and the number of paths as $\aleph=100 $ for each question node in the graph.
The embedding dimension $d$ of the skill and question representations was set to 128. 
The final dimension of the convolutional question representation was $d^{'}=256$. The size of the hidden layers of the LSTM was set to 256. The other hyperparameters were set through grid searching. The embedding matrix of the correctness and all parameter matrices in the networks were randomly initialized and updated through the training process. 

\vspace{-3 pt}
The model was optimized using Adam optimization of the learning rate on a case-by-case basis in the three datasets. The norm clipping threshold and batch size were maintained at 10 and 64, respectively. Similarly to the existing models, the sequence length of the model input was fixed at 200. 
The proposed model was implemented using TensorFlow. 
The other baselines were implemented with their best parameter settings, as specified in the original works.  
As the evaluation metric, we selected the area under the receiver operating characteristic (ROC) curve (AUC), which is widely used in existing studies. The larger the AUC value,
the better is the model performance. 

\section{Results and Analysis}

\subsection{Performance Prediction}
The different models were evaluated by their performances in predicting the future learner scores from the estimated knowledge state. Table \ref{AUC_comparison} presents the AUC results of all models on the three datasets.

\begin{table}[t]
\renewcommand{\arraystretch}{1.0}
\centering
\caption{Comparisons of the AUC results of different models on the three datasets}
\label{AUC_comparison}
\resizebox{1\linewidth}{!}{%
\begin{tabular}{ccccc}
\hline
 & {Model}  & {ASSIST0910}                        & {ASSIST1213}                         & {EdNet} \\ 
{\bf Skill-based Model}   & BKT    & 0.6571  & 0.6204   & 0.6027  \\
& DKT    & 0.7412   & 0.7256  & 0.6889    \\        & DKVMN  & 0.7559   & 0.7247 & 0.6921 \\  \hline
{\bf Question-based Model} & KTM    & 0.7582                           & 0.7212                           & 0.6899                           \\
& DKT-Q  & 0.7306                           & 0.7158                           & 0.6812                           \\
& DKT-Q\&S & 0.7616                           & 0.7389                           & 0.7235                           \\
 & DKT-CQE & 0.7998                           & 0.7686                           & 0.7523                           \\
 & GIKT   & 0.7845                           & 0.7712                           & 0.7529                           \\ 
 & DAS3H &0.789 & 0.741 & 0.731 \\
 & RKTM &0.7650 & 0.7690 & -- \\
 & AKT & 0.8152 & 0.7720 & 0. 7658 \\
& PQRLKA  & \textbf{0.8242*} & \textbf{0.7851*} & \textbf{0.7754*} \\ \hline
\end{tabular}}
\end{table}

\vspace{-3 pt}
Our model outperformed the other models on all three datasets. Specifically, the AUC scores of the PQRLKA model were 0.8242, 0.7851, and 0.7754 on the Assist0910, Assist1213 and EdNet datasets, respectively, 3.97\%, 1.39\%, and 2.25\%, respectively, above those of the state-of-the-art AKT model. Similarly to the original DKT model, our model processes the time-series data using a recurrent neural network framework, but achieved 8.3\%, 5.95\% ,and 8.65\% higher AUCs than the DKT model on the Assist0910, Assist1213 and EdNet datasets, respectively.

The skill-based BKT model was the worst performer among the models because it tracks the mastery of each skill separately, without considering a contextual trial sequence of all skills. DKT and DKVMN achieved similar performances and considerably outperformed the BKT model, confirming the effectiveness of applying deep neural networks to this task; however, they were slightly outperformed by the other question-based models. KTM framework, which incorporates several traditional models, typically obtained similar AUC scores to those of DKT and DKVMN. DKT extended with various input-question embeddings (DKT-Q, DKT-Q\&S, and DKT-CQE) demonstrated noticeable performance differences. DKT-Q using the one-hot encoding of question representations performed much worse than the original DKT model, owing to the sparsity of question interactions in these datasets. DKT-Q\&S and DKT-CQE decidedly outperformed the original DKT and DKT-Q models, consistent with our intuition that each question contains distinctive information even when it requires the same skills as one or more other questions in the dataset. Therefore, incorporating the distinctive question and skill information into the question representations can improve the model performance. The comparison between DKT and DKT-CQE also shows the effectiveness of the proposed convolutional question representation. Moreover, the improvements in the AUC scores of these DKT-extended models were smaller on Assist1213 than on the other two datasets. As Assist1213 is a single-skill dataset, incorporating the skill information into the question representation provides less additional information on this dataset than on the other datasets. 
DAS3H and RKTM are built on the KTM model by considering additional rich information, and show better results than the original KTM model. Here we do not present the results of RKTM on the EdNet as the code for implementing RKTM is not published by the author.      
Our PQRLKA model outperformes GIKT and AKT, which also take the well-designed question embedding as input. This result validates the effectiveness of the Q-matrix refinement for the task of LKA.

\subsection{Model Analysis}

\subsubsection{Comparison of Prerequisite Inference by Different Methods} 

In subsection 3.1.1, we presented eight methods for inferring prerequisites to build the KG from the response data. This subsection compares the performances of these methods.

\begin{figure}[t]
\centering
\includegraphics[width=0.8\linewidth]{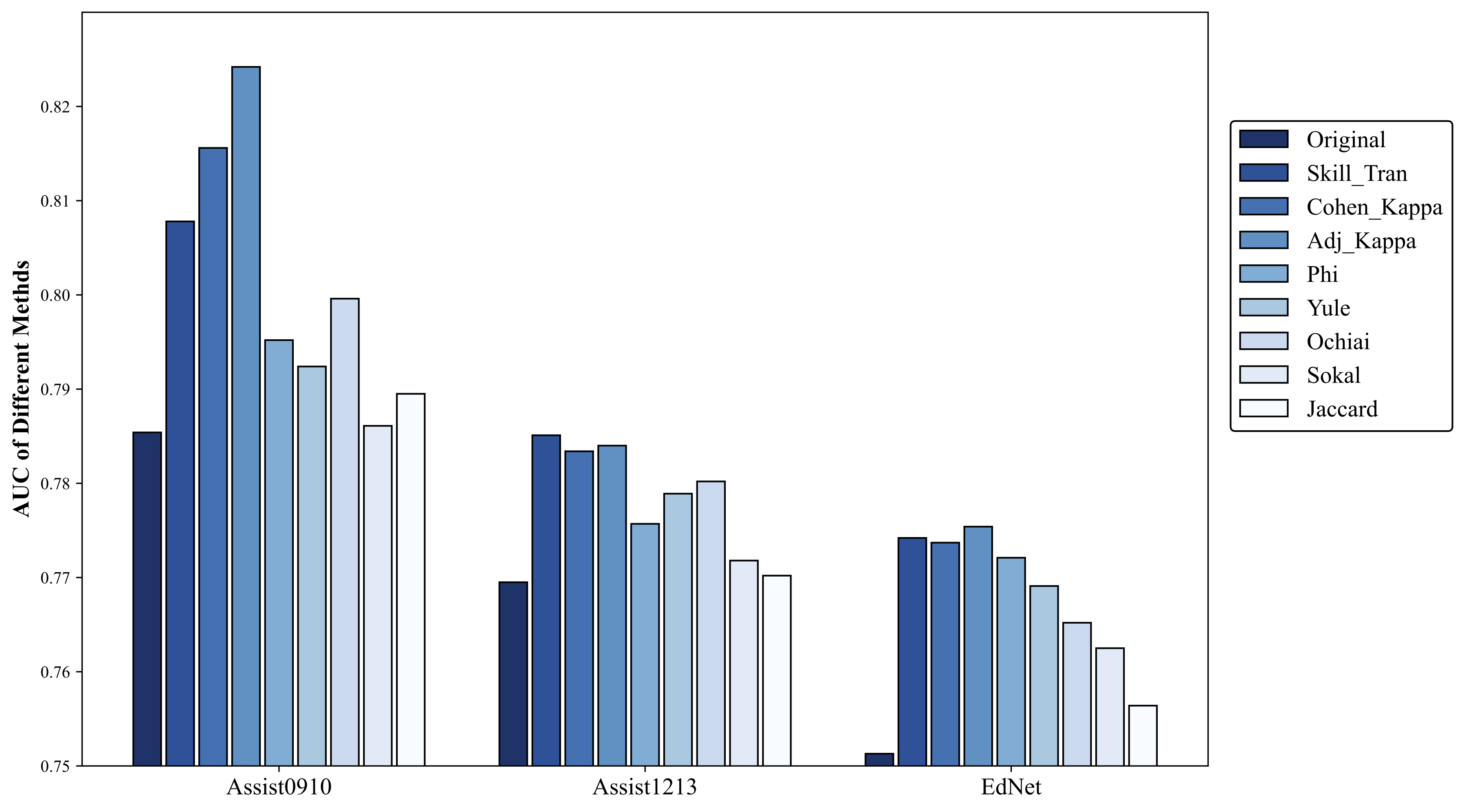} 
\caption{KS inferred from the learner response data}
\label{AUC_diffKS}
\end{figure}

Figure \ref{AUC_diffKS} compares the AUC results of the predictions of nine methods on the three datasets. Here the ``original'' method represents the embedding learning on the original question--skill relation graph and the other eight methods are based on the prerequisite-enhanced graph. As evidenced in the figure, the eight KG enhanced methods generally outperformed the ``original'' method on all three datasets, validating the effectiveness of the KG enhanced graph in KT tasks. The adjusted Kappa yielded the best performance on both Assist0910 and EdNet, whereas the skill transaction method performed best on Assist1213.

\vspace{-5 pt}
\subsubsection{Visualization of Inferred Prerequisites} 

\begin{figure}[tbp]
\centering
\includegraphics[width=1\linewidth]{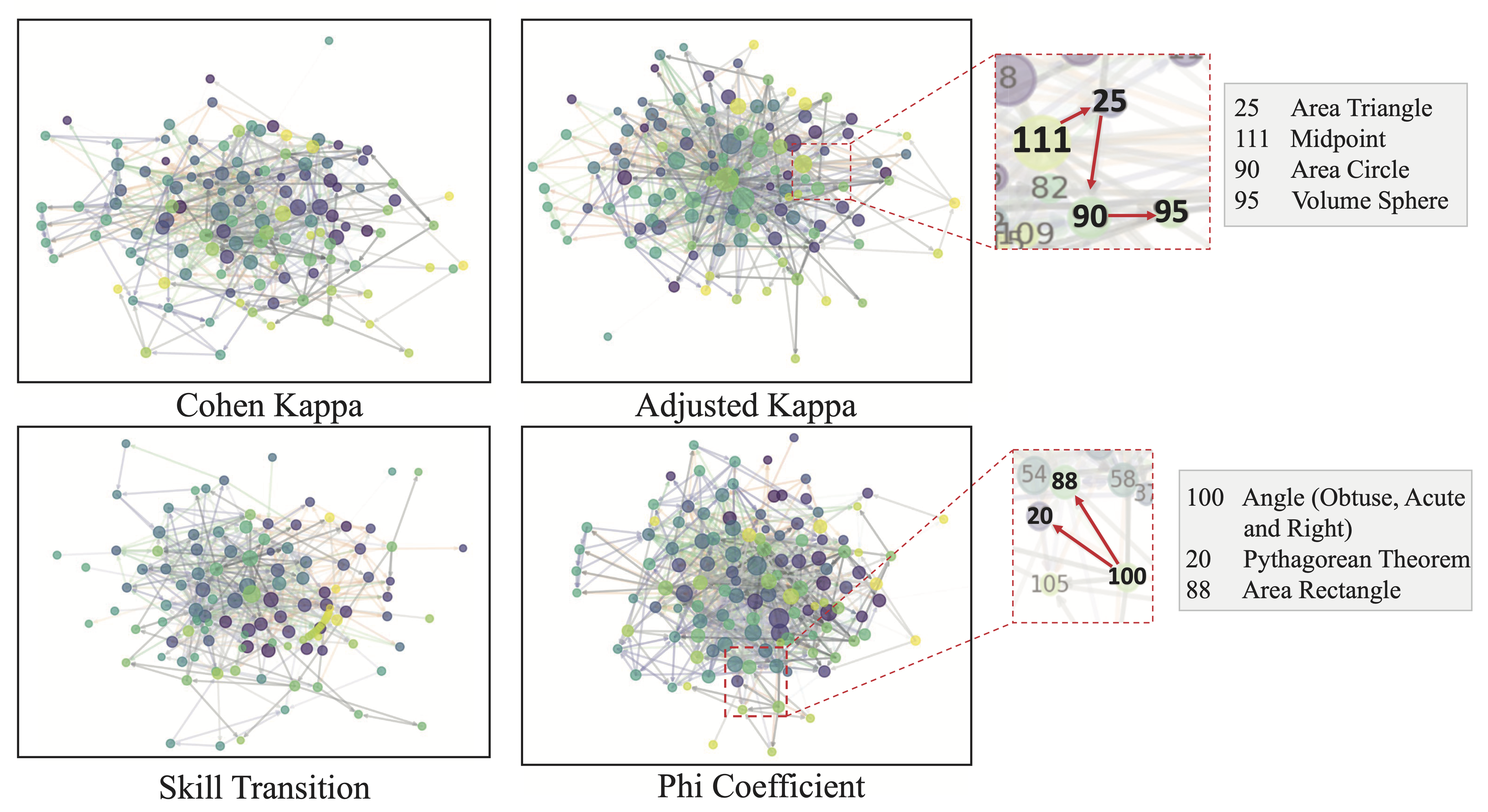} 
\caption{Visualization of the directed KG graphs generated by four methods on Assist0910}
\label{KS_show}
\end{figure}

Figure \ref{KS_show} illustrates the KGs generated from the prerequisites using four methods, inferred from the learner response data in Assist0910. The right-hand side of this figure enlarges a part of the graphs to show their local connections. The nodes and edges in the KG form dense graphs with 
similar structures, showing several interconnected nodes. These graphs also show some interesting properties. In the adjust-Kappa graph, four nodes (25, 111, 90, and 95) were locally interconnected and revealed a perfect ordering of the skills (prerequisite and post-requisite relations) in the geometry. The local connections in the Phi coefficient graph also presented reasonable relations among the three skills. These results confirm that our KG discovery methods can infer prerequisite skill pairs from the ordering of learners' mastery of skills. 

\section{Conclusions}
In this paper we proposed a prerequisite-driven Q-matrix refinement framework for learner knowledge assessment in online learning context. We first explored eight methods to infer the prerequisites from learners' response data in the online learning systems and used it to refine the expert-defined Q-matrix to eliminate the potential subjective tendency of experts in designing the Q-matrix. This refinement leveraged the additional structural information (prerequisites) to 
allow the interpretability of the optimized Q-matrix while enabling the scalability to apply to the large-scale online learning context. 
Based on the refined Q-matrix, we proposed a Metapath2Vec enhanced convolutional representation method to obtain the comprehensive representations of the attempted items with rich information. 
These representations for the learners' exercising sequences are fed into the PQRLKA model, which considers the long-term dependencies using an attention mechanism, to finally predict the learners' performance on new items. Extensive experiments conducted on three real-world datasets demonstrated the capability and interpretability of our model to infer the prerequisites from the learning data, and the better performance of the embedding representation on the prerequisite enhanced graph, thus contributing to the superiority of the proposed model for learner knowledge assessment, and validating its potential applicability to real online learning environments. 

\vspace{-5 pt}
In future work, we intend to embed KG learning into our model in an end-to-end manner. The KG will then be automatically learned in the training process rather than computed from the learner performance data. Such automation can potentially expand the KG to new skills in the domain, for example, enlarging the KG of mathematics from the primary school to high school level. 



%


\begin{thebibliography}{10}

\bibitem{baker2001basics}
F.~B. Baker.
\newblock {\em The basics of item response theory}.
\newblock ERIC, 2001.

\bibitem{cen2006learning}
H.~Cen, K.~Koedinger, and B.~Junker.
\newblock Learning factors analysis--a general method for cognitive model
  evaluation and improvement.
\newblock In {\em ITS}, pages 164--175. Springer, 2006.

\bibitem{chen2017q}
C.~Chen.
\newblock {\em Q-matrix optimization for cognitive diagnostic assessment}.
\newblock PhD thesis, University of Illinois at Urbana-Champaign, 2017.

\bibitem{chen2018prerequisite}
P.~Chen, Y.~Lu, V.~W. Zheng, and Y.~Pian.
\newblock Prerequisite-driven deep knowledge tracing.
\newblock In {\em 2018 IEEE International Conference on Data Mining (ICDM)},
  pages 39--48. IEEE, 2018.

\bibitem{chiu2013statistical}
C.-Y. Chiu.
\newblock Statistical refinement of the q-matrix in cognitive diagnosis.
\newblock {\em Applied Psychological Measurement}, 37(8):598--618, 2013.

\bibitem{choffin2019das3h}
B.~Choffin, F.~Popineau, Y.~Bourda, and J.-J. Vie.
\newblock Das3h: Modeling student learning and forgetting for optimally
  scheduling distributed practice of skills.
\newblock {\em Proceedings of the 12th EDM}, 2019.

\bibitem{choi2020ednet}
Y.~Choi, Y.~Lee, D.~Shin, J.~Cho, S.~Park, S.~Lee, J.~Baek, C.~Bae, B.~Kim, and
  J.~Heo.
\newblock Ednet: A large-scale hierarchical dataset in education.
\newblock In {\em International Conference on Artificial Intelligence in
  Education}, pages 69--73. Springer, 2020.

\bibitem{corbett1994knowledge}
A.~T. Corbett and J.~R. Anderson.
\newblock Knowledge tracing: Modeling the acquisition of procedural knowledge.
\newblock {\em User modeling and user-adapted interaction}, 4(4):253--278,
  1994.

\bibitem{de2008empirically}
J.~De~La~Torre.
\newblock An empirically based method of q-matrix validation for the dina
  model: Development and applications.
\newblock {\em Journal of educational measurement}, 45(4):343--362, 2008.

\bibitem{de2009dina}
J.~De~La~Torre.
\newblock Dina model and parameter estimation: A didactic.
\newblock {\em Journal of educational and behavioral statistics},
  34(1):115--130, 2009.

\bibitem{de2016general}
J.~de~la Torre and C.-Y. Chiu.
\newblock A general method of empirical q-matrix validation.
\newblock {\em Psychometrika}, 81(2):253--273, 2016.

\bibitem{desmarais2014refinement}
M.~Desmarais, B.~Beheshti, and P.~Xu.
\newblock The refinement of a q-matrix: Assessing methods to validate tasks to
  skills mapping.
\newblock In {\em Educational Data Mining 2014}, 2014.

\bibitem{desmarais2012mapping}
M.~C. Desmarais.
\newblock Mapping question items to skills with non-negative matrix
  factorization.
\newblock {\em ACM SIGKDD Explorations Newsletter}, 13(2):30--36, 2012.

\bibitem{desmarais2013matrix}
M.~C. Desmarais and R.~Naceur.
\newblock A matrix factorization method for mapping items to skills and for
  enhancing expert-based q-matrices.
\newblock In {\em International Conference on Artificial Intelligence in
  Education}, pages 441--450. Springer, 2013.

\bibitem{desmarais2015combining}
M.~C. Desmarais, P.~Xu, and B.~Beheshti.
\newblock Combining techniques to refine item to skills q-matrices with a
  partition tree.
\newblock {\em International Educational Data Mining Society}, 2015.

\bibitem{dong2017metapath2vec}
Y.~Dong, N.~V. Chawla, and A.~Swami.
\newblock metapath2vec: Scalable representation learning for heterogeneous
  networks.
\newblock In {\em Proceedings of the 23rd ACM SIGKDD international conference
  on knowledge discovery and data mining}, pages 135--144, 2017.

\bibitem{feng2009addressing}
M.~Feng, N.~Heffernan, and K.~Koedinger.
\newblock Addressing the assessment challenge with an online system that tutors
  as it assesses.
\newblock {\em User Modeling and User-Adapted Interaction}, 19(3):243--266,
  2009.

\bibitem{Gan2020AIModeling}
W.~Gan, Y.~Sun, X.~Peng, and Y.~Sun.
\newblock Modeling learner’s dynamic knowledge construction procedure and
  cognitive item difficulty for knowledge tracing.
\newblock {\em Applied Intelligence}, 50(11):3894--3912, 2020.

\bibitem{gan2020knowledge}
W.~Gan, Y.~Sun, and Y.~Sun.
\newblock Knowledge interaction enhanced knowledge tracing for learner
  performance prediction.
\newblock In {\em 2020 7th International Conference on Behavioural and Social
  Computing (BESC)}, pages 1--6. IEEE, 2020.

\bibitem{gan2022knowledge}
W.~Gan, Y.~Sun, and Y.~Sun.
\newblock Knowledge interaction enhanced sequential modeling for interpretable
  learner knowledge diagnosis in intelligent education systems.
\newblock {\em Neurocomputing}, 488:36--53, 2022.

\bibitem{gan2022knowledgestructure}
W.~Gan, Y.~Sun, and Y.~Sun.
\newblock Knowledge structure enhanced graph representation learning model for
  attentive knowledge tracing.
\newblock {\em International Journal of Intelligent Systems}, 37(3):2012--2045,
  2022.

\bibitem{gan2019field}
W.~Gan, Y.~Sun, S.~Ye, Y.~Fan, and Y.~Sun.
\newblock Field-aware knowledge tracing machine by modelling students' dynamic
  learning procedure and item difficulty.
\newblock In {\em 2019 International Conference on Data Mining Workshops
  (ICDMW)}, pages 1045--1046. IEEE, 2019.

\bibitem{gan2019ai}
W.~Gan, Y.~Sun, S.~Ye, Y.~Fan, and Y.~Sun.
\newblock \protect{AI-Tutor}: Generating tailored remedial questions and
  answers based on cognitive diagnostic assessment.
\newblock In {\em 2019 6th International Conference on Behavioral, Economic and
  Socio-Cultural Computing (BESC)}, pages 1--6. IEEE, 2019.

\bibitem{gasparetti2022discovering}
F.~Gasparetti.
\newblock Discovering prerequisite relations from educational documents through
  word embeddings.
\newblock {\em Future Generation Computer Systems}, 127:31--41, 2022.

\bibitem{ghosh2020context}
A.~Ghosh, N.~Heffernan, and A.~S. Lan.
\newblock Context-aware attentive knowledge tracing.
\newblock In {\em Proceedings of the 26th ACM SIGKDD International Conference
  on Knowledge Discovery \& Data Mining}, pages 2330--2339, 2020.

\bibitem{hasanov2019survey}
A.~Hasanov, T.~H. Laine, and T.-S. Chung.
\newblock A survey of adaptive context-aware learning environments.
\newblock {\em Journal of Ambient Intelligence and Smart Environments},
  11(5):403--428, 2019.

\bibitem{huang2020learning}
Z.~Huang, Q.~Liu, Y.~Chen, L.~Wu, K.~Xiao, E.~Chen, H.~Ma, and G.~Hu.
\newblock Learning or forgetting? a dynamic approach for tracking the knowledge
  proficiency of students.
\newblock {\em ACM Transactions on Information Systems (TOIS)}, 38(2):1--33,
  2020.

\bibitem{jiangimproving2021}
C.~Jiang, W.~Gan, G.~Su, Y.~Sun, and Y.~Sun.
\newblock Improving knowledge tracing through embedding based on metapath.
\newblock In {\em The 29th International Conference on Computers in Education
  (ICCE)}, pages 11--20. Asia-Pacific Society for Computers in Education, 2021.

\bibitem{kang2019q}
C.~Kang, Y.~Yang, and P.~Zeng.
\newblock Q-matrix refinement based on item fit statistic rmsea.
\newblock {\em Applied psychological measurement}, 43(7):527--542, 2019.

\bibitem{lai2021recurrent}
Z.~Lai, L.~Wang, and Q.~Ling.
\newblock Recurrent knowledge tracing machine based on the knowledge state of
  students.
\newblock {\em Expert Systems}, page e12782, 2021.

\bibitem{liu2012data}
J.~Liu, G.~Xu, and Z.~Ying.
\newblock Data-driven learning of q-matrix.
\newblock {\em Applied psychological measurement}, 36(7):548--564, 2012.

\bibitem{liu2021survey}
Q.~Liu, S.~Shen, Z.~Huang, E.~Chen, and Y.~Zheng.
\newblock A survey of knowledge tracing.
\newblock {\em arXiv preprint arXiv:2105.15106}, 2021.

\bibitem{liu2019exploiting}
Q.~Liu, S.~Tong, C.~Liu, H.~Zhao, E.~Chen, H.~Ma, and S.~Wang.
\newblock Exploiting cognitive structure for adaptive learning.
\newblock In {\em Proceedings of the 25th ACM SIGKDD International Conference
  on Knowledge Discovery \& Data Mining}, pages 627--635, 2019.

\bibitem{liu2020improving}
Y.~Liu, Y.~Yang, X.~Chen, J.~Shen, H.~Zhang, and Y.~Yu.
\newblock Improving knowledge tracing via pre-training question embeddings.
\newblock {\em arXiv preprint arXiv:2012.05031}, 2020.

\bibitem{ma2020empirical}
W.~Ma and J.~de~la Torre.
\newblock An empirical q-matrix validation method for the sequential
  generalized dina model.
\newblock {\em British Journal of Mathematical and Statistical Psychology},
  73(1):142--163, 2020.

\bibitem{matsuda2015machine}
N.~Matsuda, T.~Furukawa, N.~Bier, and C.~Faloutsos.
\newblock Machine beats experts: Automatic discovery of skill models for
  data-driven online course refinement.
\newblock {\em International Educational Data Mining Society}, 2015.

\bibitem{minn2016refinement}
S.~Minn, M.~C. Desmarais, and S.~Fu.
\newblock Refinement of a q-matrix with an ensemble technique based on
  multi-label classification algorithms.
\newblock In {\em European Conference on Technology Enhanced Learning}, pages
  165--178. Springer, 2016.

\bibitem{nakagawa2019graph}
H.~Nakagawa, Y.~Iwasawa, and Y.~Matsuo.
\newblock Graph-based knowledge tracing: modeling student proficiency using
  graph neural network.
\newblock In {\em 2019 IEEE/WIC/ACM International Conference on Web
  Intelligence (WI)}, pages 156--163. IEEE, 2019.

\bibitem{nazaretsky2019kappa}
T.~Nazaretsky, S.~Hershkovitz, and G.~Alexandron.
\newblock Kappa learning: A new item-similarity method for clustering
  educational items from response data.
\newblock {\em International Educational Data Mining Society}, 2019.

\bibitem{pan2017prerequisite}
L.~Pan, C.~Li, J.~Li, and J.~Tang.
\newblock Prerequisite relation learning for concepts in moocs.
\newblock In {\em Proceedings of the 55th Annual Meeting of the Association for
  Computational Linguistics (Volume 1: Long Papers)}, pages 1447--1456, 2017.

\bibitem{pandey2020rkt}
S.~Pandey and J.~Srivastava.
\newblock Rkt: Relation-aware self-attention for knowledge tracing.
\newblock In {\em Proceedings of the 29th ACM International Conference on
  Information \& Knowledge Management}, pages 1205--1214, 2020.

\bibitem{pavlik2009performance}
P.~I. Pavlik~Jr, H.~Cen, and K.~R. Koedinger.
\newblock Performance factors analysis--a new alternative to knowledge tracing.
\newblock {\em Online Submission}, 2009.

\bibitem{pelanek2019measuring}
R.~Pel{\'a}nek.
\newblock Measuring similarity of educational items: An overview.
\newblock {\em IEEE Transactions on Learning Technologies}, 13(2):354--366,
  2019.

\bibitem{piech2015deep}
C.~Piech, J.~Bassen, J.~Huang, S.~Ganguli, M.~Sahami, L.~J. Guibas, and
  J.~Sohl-Dickstein.
\newblock Deep knowledge tracing.
\newblock In {\em Advances in neural information processing systems}, pages
  505--513, 2015.

\bibitem{sabnis2021upreg}
V.~Sabnis, K.~Abhinav, V.~Subramanian, A.~Dubey, and P.~Bhat.
\newblock Upreg: An unsupervised approach for building the concept prerequisite
  graph.
\newblock {\em International Educational Data Mining Society}, 2021.

\bibitem{shin2020saint+}
D.~Shin, Y.~Shim, H.~Yu, S.~Lee, B.~Kim, and Y.~Choi.
\newblock Saint+: Integrating temporal features for ednet correctness
  prediction.
\newblock {\em arXiv preprint arXiv:2010.12042}, 2020.

\bibitem{sun2014alternating}
Y.~Sun, S.~Ye, S.~Inoue, and Y.~Sun.
\newblock Alternating recursive method for q-matrix learning.
\newblock In {\em Proceedings of the 7th International Conference on
  Educational Data Mining}, pages 14--20, 2014.

\bibitem{tatsuoka2013toward}
K.~K. Tatsuoka.
\newblock {\em Toward an Integration of IteD1-Response Theory and Cognitive
  Error Diagnosis}.
\newblock Routledge, 2013.

\bibitem{tong2020hgkt}
H.~Tong, Y.~Zhou, and Z.~Wang.
\newblock Hgkt: Introducing problem schema with hierarchical exercise graph for
  knowledge tracing.
\newblock {\em arXiv preprint arXiv:2006.16915}, 2020.

\bibitem{vaswani2017attention}
A.~Vaswani, N.~Shazeer, N.~Parmar, J.~Uszkoreit, L.~Jones, A.~N. Gomez,
  L.~Kaiser, and I.~Polosukhin.
\newblock Attention is all you need.
\newblock {\em arXiv preprint arXiv:1706.03762}, 2017.

\bibitem{vie2019knowledge}
J.-J. Vie and H.~Kashima.
\newblock Knowledge tracing machines: Factorization machines for knowledge
  tracing.
\newblock In {\em Proceedings of the AAAI Conference on Artificial
  Intelligence}, volume~33, pages 750--757, 2019.

\bibitem{wang2021learning}
C.~Wang and J.~Lu.
\newblock Learning attribute hierarchies from data: Two exploratory approaches.
\newblock {\em Journal of Educational and Behavioral Statistics}, 46(1):58--84,
  2021.

\bibitem{xu2016boosted}
P.~Xu and M.~C. Desmarais.
\newblock Boosted decision tree for q-matrix refinement.
\newblock {\em International Educational Data Mining Society}, 2016.

\bibitem{yang2020deep}
S.~Yang, M.~Zhu, J.~Hou, and X.~Lu.
\newblock Deep knowledge tracing with convolutions.
\newblock {\em arXiv preprint arXiv:2008.01169}, 2020.

\bibitem{yang2020gikt}
Y.~Yang, J.~Shen, Y.~Qu, Y.~Liu, K.~Wang, Y.~Zhu, W.~Zhang, and Y.~Yu.
\newblock Gikt: A graph-based interaction model for knowledge tracing.
\newblock {\em arXiv preprint arXiv:2009.05991}, 2020.

\bibitem{zhang2016topological}
J.~Zhang and I.~King.
\newblock Topological order discovery via deep knowledge tracing.
\newblock In {\em International Conference on Neural Information Processing},
  pages 112--119. Springer, 2016.

\bibitem{zhang2017dynamic}
J.~Zhang, X.~Shi, I.~King, and D.-Y. Yeung.
\newblock Dynamic key-value memory networks for knowledge tracing.
\newblock In {\em Proceedings of the 26th WWW}, pages 765--774, 2017.

\end{thebibliography}
%

\balancecolumns
\end{document}